\documentclass[english,aps,pra, twocolumn, floatfix, showpacs, superscriptaddress, notitlepage]{revtex4-1}
\usepackage{amsmath}
\usepackage{amsfonts}
\usepackage{graphicx}
\usepackage{color}
\usepackage{hyperref}
\usepackage{epstopdf}

 \begin{abstract}
Strong synthetic magnetic fields have been successfully implemented in periodically driven optical lattices. However, the interplay of the driving and interactions introduces detrimental heating, and for this reason it is still challenging to reach a fractional quantum Hall state in cold-atom setup. 
By performing a numerical study, we investigate stability of a bosonic Laughlin state in a small atomic sample exposed to driving.
We identify an optimal regime of microscopic parameters, in particular interaction strength $U$ and the driving frequency $\omega$, such that the stroboscopic dynamics supports the basic $\nu = 1/2$ Laughlin state. Moreover, we explore slow ramping of a driving term and show that the considered protocol allows for the preparation of the Laughlin state on experimentally realistic time scales.
\end{abstract}

\begin{document}
 \title{Bosonic fractional quantum Hall states in driven optical lattices}

\author{Ana Hudomal}
\affiliation{Scientific Computing Laboratory, Center for the Study of Complex Systems,
Institute of Physics Belgrade, University of Belgrade, Pregrevica 118, 11080 Belgrade, Serbia}
\author{Nicolas Regnault}
\affiliation{Joseph Henry Laboratories and Department of Physics, Princeton University, Princeton, New Jersey 08544, USA}
\affiliation{Laboratoire de Physique de l'\'Ecole normale sup\'erieure, ENS, Universit\'e PSL, CNRS, Sorbonne Universit\'e, Universit\'e Paris-Diderot, Sorbonne Paris Cit\'e, Paris, France}
 \author{Ivana Vasi\'c} 
\affiliation{Scientific Computing Laboratory, Center for the Study of Complex Systems,
Institute of Physics Belgrade, University of Belgrade, Pregrevica 118, 11080 Belgrade, Serbia}
 \maketitle
 
 \section{Introduction}
 
Cold atoms in optical lattices provide a highly tunable platform for quantum simulations of relevant many-body Hamiltonians  \cite{Bloch2008, Hofstetter2018}.
Since early experiments with quantum gases, there has been a strong interest in the realization of fractional quantum Hall (FQH) states in these setups \cite{Wilkin2000, Cooper2001, Paredes2001, Popp2004, Sorensen, Rezayi2005, Hafezi, Petrescu2017, He2017, Rosson2019}. Despite numerous experimental achievements and a variety of theoretical proposals, FQH physics has still not been reached in cold-atom experiments.

A milestone in the field has been recently achieved by the realization of artificial gauge potentials \cite{Lin2009, Miyake2013, Aidelsburger2013, Jotzu, Aidelsburger2015, Kennedy2015, Flaschner2016, Tai2017, Dalibard2011, Eckardt2017, Cooper2019}. In particular, the topological index of a resulting energy band of an optical lattice featuring a strong synthetic magnetic field has been directly probed \cite{Aidelsburger2015}. At first glance, both key requirements for the emergence of FQH states - atomic interactions and strong synthetic magnetic fields -   are now experimentally available. However, there are several specific details in the implementation of strong  synthetic magnetic fields for cold atoms that make the realization of FQH states still challenging.

 The most advanced recent realizations of artificial gauge potentials exploit periodically driven optical lattices \cite{ Miyake2013, Aidelsburger2013, Jotzu, Aidelsburger2015, Kennedy2015, Flaschner2016, Tai2017, Dalibard2011, Eckardt2017, Cooper2019}. Using Floquet theory, the stroboscopic dynamics of a non-interacting driven system can be related to an effective time-independent Hamiltonian \cite{Goldman2014, Goldman2015, Eckardt2015, Plekhanov2017}.
 This approach - Floquet engineering -  enriches the set of quantum models that can be simulated in cold-atom experiments. However, general arguments and numerical studies \cite{DAlessio2014, Lazarides2014, Ponte2015} suggest that the interplay of interactions and driving in a thermodynamically large system introduces heating, leading to a featureless infinite-temperature state in the long-time limit. 
 
 Although this general result might sound discouraging, the heating process can be very slow in some driven systems for specific regime of microscopic parameters. There, the system can be described by a physically interesting ``prethermal'' Floquet state on experimentally relevant time scales \cite{Abanin2015, Mori2016, Bukov2015, Kuwahara2016, Abanin2017, Abanin2017p2, Machado2017}. Moreover, the onset of thermalization in a finite-size interacting system may exhibit unexpected features, not found in the thermodynamic limit \cite{Haldar2018, Seetharam2018}.
 Heating rates and resulting instabilities have been recently investigated both theoretically and experimentally for the driven Bose-Hubbard model in the weakly interacting regime \cite{Bukov2015, Lellouch2017, Nager2018, Boulier2019}. 
 
 In this paper, we consider small systems of several interacting bosonic atoms in a periodically driven optical lattice featuring synthetic magnetic flux. The focus of our study is on finding optimal microscopic parameters that would allow to prepare and probe the basic bosonic Laughlin state in this setup. 
 To this end, we employ exact numerical simulations of the driven Bose-Hubbard model \cite{Bukov2014} for small system sizes.
 
 From one point of view, it is expected that a small driven system exhibits low heating rates for a driving frequency set above  a finite bandwidth of an effective model \cite{DAlessio2014}. However, driving a system with such a high frequency may lead to undesirable effects, such as coupling of the lowest band to higher bands of the underlying optical lattice, thus making the initial description based on the lowest-band Hubbard model inapplicable. These effects have been addressed in a recent study \cite{Sun2018} where an optimal intermediate frequency window for Floquet engineering has been established. 
 
 In our study, we go a step further in the search for the optimal regime that might allow for the bosonic Laughlin states under driving. In particular, for a realistic, intermediate value of a driving frequency, the interaction term complicates the effective model by introducing several higher-order terms. Their effect on the topological states has been addressed only recently \cite{Grushin, Raciunas} and it has been found that  typically these terms work against the topological state. For this reason, the stability of the Laughlin state at intermediate driving frequency requires a separate study, that we perform here. Moreover, we numerically investigate an experimentally relevant preparation protocol for the Laughlin state in a driven system \cite{Dauphin2017}. For a reference, we note that a simpler, but closely related question concerning the static (undriven systems) has gained lot of attention \cite{Popp2004, Sorensen, Motruk2017, He2017}.
 
 The paper is organized as follows: in Section \ref{sec:Model} we introduce the model under study and briefly review key features of the particle-entanglement spectra that we will exploit in the identification of the Laughlin-like state. Then, in Subsection \ref{subsec:Heating} we investigate general heating effects of interacting bosons exposed to the driving. By extending this approach, in Subsection \ref{subsec:Uf} we construct the stroboscopic time-evolution operator and inspect its eigenstates in order to identify possible FQH states. Finally, in Section \ref{sec:Slowramp} we address the possibility of accessing these states in an experiment through a slow ramp of the driving term.

 \section{Model and method}
 
 In this section we first introduce the driven model and explain the basis of Floquet engineering. Then we summarize several key features of the particle-entanglement spectra that we use to characterize the bosonic Laughlin states.
 \label{sec:Model}
  \begin{figure}[!t]
\includegraphics[width=0.96\linewidth]{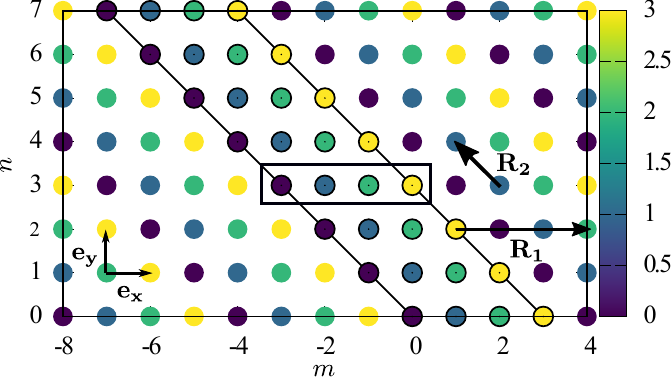}
\caption{Lattice geometry used throughout the paper.  
The parallelogram gives the exemplary lattice size $(L_x, L_y) = (4, 8)$.  The color scale is defined by $\mod(m+n, 4)$, in accordance with the driving term from Eq.~\ref{eq:Hoft}. The vectors $\mathbf{R}_1 = 4\, \mathbf{e}_x, \mathbf{R}_2 =- \mathbf{e}_x+\mathbf{e}_y$ are used to implement periodic boundary conditions.
The small rectangle gives the magnetic unit cell for the effective model in Eq.~(\ref{eq:heff}). }
\label{Fig:Fig0}
\end{figure}

\subsection{Driven model}
 Properties of bosonic atoms in a deep optical lattice can be realistically described within the framework of the Bose-Hubbard model \cite{Bloch2008}.
 We consider a basic driving scheme \cite{Bukov2014} that introduces a uniform, synthetic magnetic flux into a \emph{square} optical lattice here spanned by the two vectors ${\bf e_x}$ and ${\bf e_y}$.
 The corresponding Hamiltonian is given by the driven Bose-Hubbard model
 \begin{eqnarray}
  \hat H(t) &=& -J_x \sum_{m, n} \left(\hat  a_{m+1, n}^{\dagger} \hat a_{m, n} + \text{h.~c.~}\right)\nonumber\\
  &-& J_y  \sum_{m, n} \left(e^{i \omega t} \hat a_{m, n+1}^{\dagger}\hat a_{m, n}  + \text{h.~c.~}\right)\nonumber\\
  &+& \frac{\kappa}{2}\sum_{m, n} \sin\left(\omega\, t - (m + n -1/2)\,\phi\right) \hat n_{m,n}\nonumber\\
  &+& \frac{U}{2}\sum_{m, n}\hat n_{m, n}(\hat n_{m, n}-1),
  \label{eq:Hoft}
 \end{eqnarray}
where operators $\hat a_{m, n}$ ($\hat a^{\dagger}_{m, n}$) annihilate (create) a boson at lattice position $(m, n)$, and local density operators are $\hat n_{m,n} = \hat a^{\dagger}_{m, n} \hat a_{m, n}$. $J_x$ and $J_y$ are tunneling amplitudes and $U$ is the on-site local repulsive interaction. We use the units where $\hbar = 1$ and the lattice constant $a = 1$. The driving scheme is defined by the driving frequency $\omega$, the driving amplitude $\kappa$ and by a phase $\phi$. 
In the following we set $\phi = \pi/2$ and $\kappa/\omega = 0.5$.  These values were recently used in an experimental realization of the Harper-Hofstadter model \cite{Aidelsburger2015}. 
The derivation of this model is briefly reviewed in Appendix. 
We assume periodic boundary conditions implemented using the vectors $\mathbf{R}_1 = 4\, \mathbf{e}_x, \mathbf{R}_2 =- \mathbf{e}_x+\mathbf{e}_y$, as presented in Fig.~\ref{Fig:Fig0}. This choice is compatible with the driving term and it allows us to exploit translational symmetry by working in the fixed quasi momentum basis.

Formally, by using the Floquet theory \cite{Grifoni1998, Goldman2014, Goldman2015}, it can be shown that the full time-evolution operator corresponding to this model is given by
\begin{equation}
 \hat U(t, t_0) = e^{-i \hat K(t)} e^{-i (t - t_0) \hat {\mathcal{H}}_{\text{eff}}} e^{i\hat K(t_0)},
 \label{eq:U}
\end{equation}
where $\hat K(t)$ is a periodic ``kick'' operator $\hat K(t) = \hat K(t + 2\pi/\omega)$ and $\hat {\mathcal H}_{\text{eff}}$ is a time-independent effective Hamiltonian. 
The full-time evolution operator is  periodic as well and consequently the (quasi) eigenenergies of $\hat {\mathcal{H}}_{\text{eff}}$ are defined up to modulo $\omega$. The last equation gives formal mapping of a periodically driven system to an effective model that captures the stroboscopic time-evolution of the model. 

In the non-interacting regime, $U = 0$, there are several well controlled approximations to obtain the effective Hamiltonian.  These techniques are the essence of Floquet engineering - an approach where the driving scheme is implemented in such a way to yield a sought-after effective model.
However, according to general analytical arguments and numerical insights, the corresponding effective model of a driven interacting many-body system in the thermodynamic limit exhibits nonphysical features \cite{DAlessio2014, Lazarides2014}. In particular, the system thermalizes and in the long-time limit its steady state is a featureless, infinite-temperature state, independent of the initial state.

Here we consider small samples of several bosonic atoms. Due to a finite spectrum bandwidth, we expect the high-frequency expansion to be relevant for a finite range of the driving frequency.
Within these assumptions, the leading-order (in $1/\omega$) effective Hamiltonian is
\begin{eqnarray}
 \hat  H_{\text{eff}} &=& -J_x \sum_{m,n}\left( \hat  a^{\dagger}_{m + 1,n}  \hat a_{m, n}+ \mathrm{h.~c.~}\right) \nonumber\\ &-&J_y'\sum_{m,n}\left( e^{i  (m+n)\phi}  \hat a^{\dagger}_{m,n + 1}  \hat a_{m, n}+\mathrm{h.~c.~}\right)\nonumber\\ &+& \frac{U}{2} \sum_{m,n}  \hat n_{m,n} \left( \hat n_{m,n}-1\right).
 \label{eq:heff}
\end{eqnarray}
 The Hamiltonian (\ref{eq:heff}) features complex hopping phases $e^{i  (m + n )\phi}$ that result in a uniform synthetic magnetic flux $\phi$ per lattice plaquette. Due to the driving, the renormalized hopping amplitude along the $y$ direction turns into 
 \begin{equation}
 J_y' \equiv \frac{\kappa}{2\omega}\sin(\phi/2)\,J_y.
 \end{equation}
 For the values $\phi = 2\pi \alpha$, where the flux density $\alpha$ is set to $\alpha = 1/4$, and $\kappa/\omega = 0.5$, the tunneling amplitude along $y$ direction in the effective model is $J_y' \approx J_y \times 0.1768$.

In a certain regime of microscopic parameters, the ground state of the model defined in Eq.~(\ref{eq:heff}) is given by the lattice version of the Laughlin state \cite{Laughlin, HaldaneRezayi, Sorensen, Hafezi, Sterdyniak2012}.  The Laughlin state is stabilized
for the filling factor $\nu = N_p/N_{\phi} = 1/2$, where $ N_{\phi} = \alpha L_x \times L_y$ is the total number of fluxes ($N_{\phi}$ being an integer) and $N_p$ is the number of bosons, and for a strong-enough repulsion $U$. Another important requirement for the Laughlin state is to avoid the strong hopping anisotropy and to keep $J_x \approx J_y'$, so we set $J_x=0.2 J_y$.  We consider system sizes $N_p = 4,5,6$ and the respective lattices sizes $(L_x, L_y) = (4, 8), (4, 10)$ and $(4, 12)$, see Fig.~\ref{Fig:Fig0}, where we expect the ground state to correspond to the $\nu = 1/2$ Laughlin state. The Hilbert space sizes for $k_x = k_y = 0$ are $\text{dim} \,\mathcal{H} = 6564, 108604$, and $1913364$ respectively. For this choice of microscopic parameters, the model ground state of Eq.~(\ref{eq:heff}) is approximately two-fold degenerate. The two ground-states are found in the sectors $k_x = 0, k_y = 0$ and $k_x = 0, k_y = \pi$. We denote them by $|\psi_{\text{LGH}}^{0, 0}\rangle$ and $|\psi_{\text{LGH}}^{0,\pi}\rangle$.

\begin{figure*}[!t]
\includegraphics[width=\linewidth]{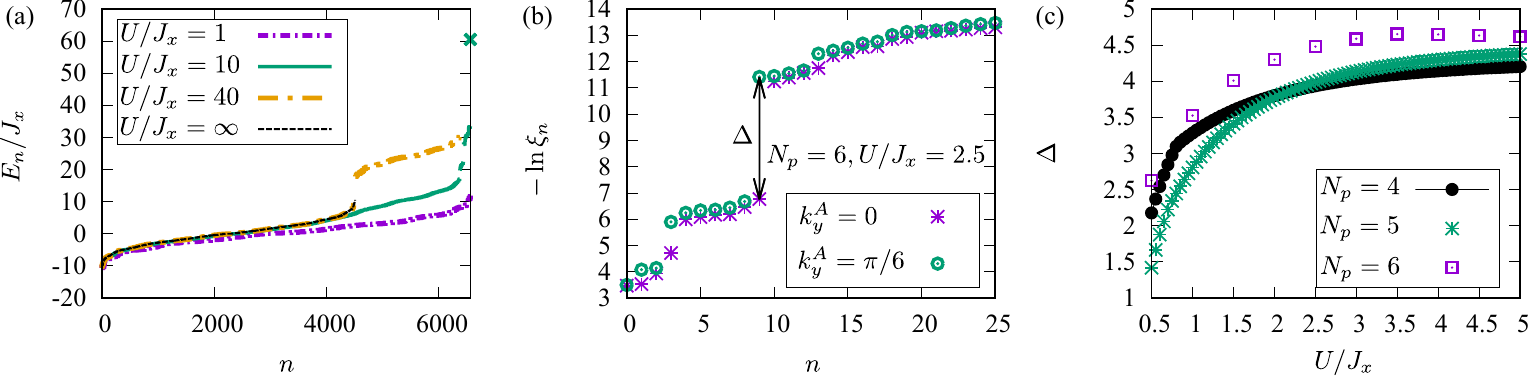}
\caption{
(a) The energy spectrum $E_n$ of the model from Eq.~(\ref{eq:heff}) in the $k_x = 0, k_y = 0$ sector  for $N_p = 4$ and different values of interaction $U/J_x = 1, 10, 40$ and $U/J_x=\infty$ (hard-core bosons). The top part of the spectrum is at  $\approx (U/J_x) N_p \left(N_p-1\right)/2$. (Not shown for $U/J_x = 40$.) For a high ratio $U /J_x$ the spectrum splits into bands. The lowest band corresponds to hard-core bosons. (b) The low-lying part of the particle-entanglement spectrum $-\ln \xi_n$ of the ground-state incoherent superposition, Eq.~(\ref{eq:rho}),  in the region $A$ momentum sectors $k_y^A=0$ and $k_y^A = \pi/6$, and for $N_p = 6, U/J_x = 2.5$. (c) The particle-entanglement gap $\Delta$ of the incoherent superposition  Eq.~(\ref{eq:rho}) as a function of interaction strength $U$ for $N_p = 4,5,6$.}
\label{Fig:Fig1}
\end{figure*}

As we are mainly interested in the driven regime, it is not only the ground state, but the full spectrum of the model from Eq.~(\ref{eq:heff}) that plays a role. A rough argument is that the system does not absorb energy provided that the driving frequency $\omega$ is set above the bandwidth of the effective model. Several spectra of the model from Eq.~(\ref{eq:heff}) for $k_x=0, k_y = 0$ are presented in Fig.~\ref{Fig:Fig1}(a). It can be seen that the ground-state energy is weakly affected by the value of $U\geq J_x$, while the top part of the spectrum with few states is found at $U N_p (N_p - 1)/2$. For higher values of $U$ the spectrum splits into bands where the lowest band corresponds to the hard-core bosons and higher bands include double and higher occupancies.

\subsection{Particle-entanglement spectra} 
There are several ways to characterize the ground-states of the model from Eq.~(\ref{eq:heff}) as the Laughlin states. Usually, the starting point in this direction is the identification of the two-fold degeneracy expected in the implemented torus geometry for $\nu = 1/2$. Another relevant quantity is the overlap of the numerically obtained state with the Laughlin analytical wave function in the torus geometry \cite{Hafezi, Sterdyniak2012}. A more direct evidence can be obtained through the calculation of the relevant topological index (Chern number) or the quantized Hall conductance. One more convincing approach, that we pursue here, is based on the analysis of the entanglement spectra of the relevant states.

In the following we will use the particle-entanglement spectrum (PES)  \cite{Sterdyniak2011, Sterdyniak2012} to distinguish possible topologically non-trivial states. In order to obtain this type of entanglement spectrum, we partition $N_p$ particles into two sets  of $N_A$ and $N_B = N_p - N_A$ particles. For a given mixed state $\rho$, we construct a reduced density matrix $\rho_A = \text{tr}_B \rho$ by performing a partial trace over $N_B$ particles. The resulting PES is given by $-\ln \xi_n$, where $\xi_n$ are eigenvalues of $\rho_A$. 
The related particle-entanglement entropy is given by \cite{Zozulya2008, Haque2009}
\begin{equation}
 S_A = -\text{tr} \left(\rho_A\, \ln \rho_A \right).
 \label{eq:sa}
\end{equation}

By partitioning particles, we keep the geometry of the system unchanged. For this reason, we will inspect the PES for the different momentum sectors $k_y^A$ of the remaining $N_A$ particles. An example of a PES is presented in Fig.~\ref{Fig:Fig1}(b). 
As proposed in Refs.~\cite{Sterdyniak2011, Sterdyniak2012}, we have considered the incoherent superposition of the almost twofold degenerate ground state of Eq.~(\ref{eq:heff}) as the density matrix
\begin{equation}
 \rho_{\text{GS}} =\frac{1}{2}\left(|\psi_{\text{LGH}}^{0, 0} \rangle \langle \psi_{\text{LGH}}^{0, 0}|+| \psi_{\text{LGH}}^{0, \pi}\rangle \langle \psi_{\text{LGH}}^{0, \pi}|\right).
 \label{eq:rho}
\end{equation}
For simplicity, we only present the PES for the two momenta $k_y^A = 0$ and $k_y^A = \pi/6$. We observe a clear particle-entanglement gap $\Delta$. In addition, the counting of low-lying modes below this gap ($10$ modes for $k_y^A = 0$ and $9$ modes for $k_y^A = \pi/6$, at $N_A = 3, N_p =6$) corresponds to the Laughlin state  \cite{Sterdyniak2011, Sterdyniak2012}. In this way the PES encodes topological features of the state $\rho$ in the form of well defined number of excitations per momentum sector $k_y^A$ \cite{Sterdyniak2011, Sterdyniak2012}. This type of analysis is useful as it can identify topological features even without model states, as done for the case of fractional Chern insulators \cite{Regnault2011, Bernevig2012}.

In the following we will consider specific particle partitions $N_A=2, N_p = 4$,  $N_A=2, N_p=5$ and $N_A=3, N_p = 6$. For these  cases  the counting of excitations $\mathcal{N}_L(k_y^A)$ per momentum sector $k_y^A$ is well established and given in Table \ref{Tab:Tab1}.
 In Fig.~\ref{Fig:Fig1}(c) we show  the particle-entanglement gap of the mixtures, Eq.~(\ref{eq:rho}), obtained at different values of $U$. Numerical results for the obtained PES indicate that a reasonably large gap is found starting at $U\sim 0.5 J_x$ and the characteristic features of the Laughlin state persist with a further increase in $U$.
We note that at lower values of the flux density $\alpha < 1/4$, the Laughlin state can be found at even lower values of the repulsion $U$, \cite{Hafezi, Sterdyniak2012}.
\begin{table}[h!]
  \begin{center}
   
    \label{tab:table1}
    \begin{tabular}{|c|c|c|c|} 
      \hline
      $N_p$ & $(L_x, L_y)$ & $N_A$ &PES: ${\mathcal{N}_L(k_y^A)}$\\
      \hline\hline
      $4$   & $(4, 8)$ & 2 & $3,2,3,2,3,2,3,2$\\
      \hline
       $5$   & $(4, 10)$ & 2 & $4,3,4,3,4,3,4,3, 4, 3$\\
      \hline
       $6$   & $(4, 12)$ & 3 & $10,9,9,10,9,9, 10,9,9, 10,9,9$\\
      \hline
    \end{tabular}
  \end{center}
   \caption{Counting of modes $\mathcal{N}_L\left(k_y^A\right)$ in the PES of the Laughlin state for several system sizes and particle partitions. The last column lists the ${\mathcal{N}_L(k_y^A)}$ values for each momentum sector $k_y^A = 2 \pi i/L_y, i = 0, \ldots, L_y-1$. }
   \label{Tab:Tab1}
\end{table}

By analyzing the effective model from Eq.~(\ref{eq:Hoft}), we have obtained a guidance for the regime of microscopic parameters and for the geometry of the small system that can give rise to Laughlin states.
In the next sections our aim is to go beyond the effective model from Eq.~(\ref{eq:heff}) and to identify topological states supported by the full driven dynamics as captured by the model given in Eq.~(\ref{eq:Hoft}).

 \begin{figure*}[!t]
\includegraphics[width=\linewidth]{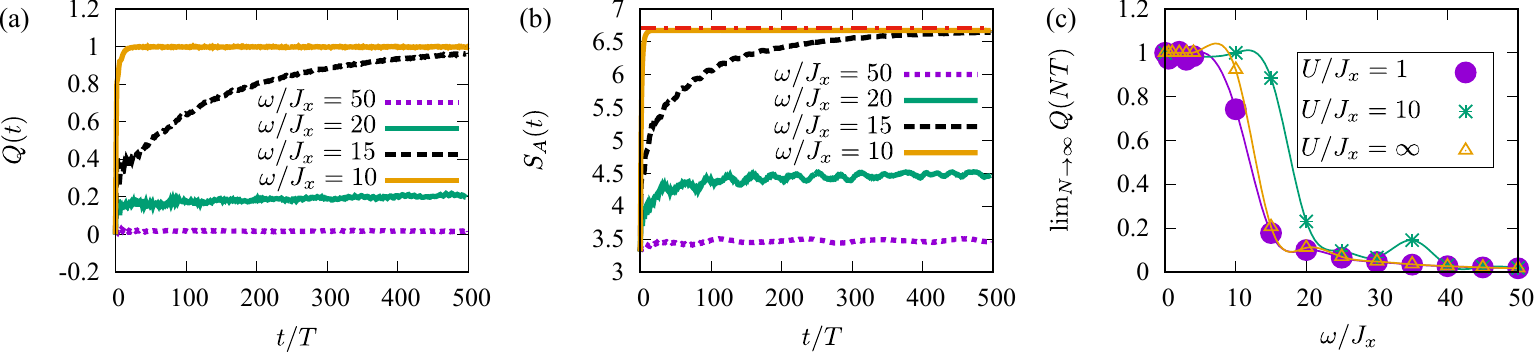}
\caption{(a) The normalized total energy $ Q(t = N T) $ from Eq.~(\ref{eq:q}), and the (b) particle-entanglement entropy $S_A(t = N T)$, Eq.~(\ref{eq:sa}),  during the time evolution governed by Eq.~(\ref{eq:Hoft}) for several driving frequencies $\omega/J_x = 50,\, 20,\, 15,\, 10$. Parameters: $N_p = 5, U/J_x = 10$. Note that the asymptotic value of $S_A$ for $\omega/J_x = 10$ and $\omega/J_x = 15$ matches the one given in Eq.~(\ref{eq:samax}), as presented by the horizontal line. (c) The long-time limit $ \lim_{N\rightarrow\infty}Q(N T)$ for $N_p = 4$ and the on-site interactions $U/J_x = 1,10$ and $U/J_x = \infty$ (hard-core bosons). The lines are only guides to the eye.  
}
\label{Fig:Fig2}
\end{figure*}
\section{Driven dynamics}
\label{sec:dynamics}
In this section we discuss the full driven dynamics as captured by the model given in Eq.~(\ref{eq:Hoft}).
\subsection{Heating}
\label{subsec:Heating}
 First we address the onset of heating following the standard procedure discussed in Refs.~\cite{Bukov2016, Machado2017}. The initial state of the system is prepared using the ground state of the effective model
\begin{equation}
 |\psi(t = 0)\rangle = e^{-i\hat K(t = 0)}|\psi_{\text{LGH}}^{0, 0}\rangle 
 \label{eq:hinitial}
\end{equation}
and we monitor the stroboscopic time-evolution $t = N \, T$, $T\equiv2\pi/\omega$ governed by the full driven model defined in Eq.~(\ref{eq:Hoft}). In our numerical simulations, we approximate the micromotion operator $\hat K(t = 0)$ using the leading-order high-frequency expansion, see Eq.~(\ref{eq:kick}).
The quantity of interest is the expectation value of the effective Hamiltonian (\ref{eq:heff})
\begin{equation}
 \langle \hat H_{\text{eff}}(t = N T) \rangle_K = \langle \psi(t)|e^{-i\hat K(t = 0)}\hat H_{\text{eff}}e^{i\hat K(t = 0)}|\psi(t)\rangle.
\label{eq:Hexp}
\end{equation}
We expect this quantity to reasonably correspond to the ground-state energy of the effective model $E_0$ in the regime of very high frequency.
On the other hand, for a ``low'' driving frequency we expect the system to quickly reach the infinite-temperature  $\beta  \rightarrow 0$ regime defined by
\begin{equation}
 \lim_{\beta  \rightarrow 0}\langle\hat  H_{\text{eff}} \rangle =\frac{1}{\text{dim} \mathcal{H}} \text{tr}\left( \hat H_{\text{eff}}\right).
\end{equation}
For this reason we monitor the normalized total energy
\begin{equation}
 Q(t = N T) = \frac{\langle \hat H_{\text{eff}}(t = N T) \rangle_K - E_0 }{\lim_{\beta  \rightarrow 0}\langle\hat  H_{\text{eff}} \rangle - E_0}
 \label{eq:q}
\end{equation}
and we present it in Fig.~\ref{Fig:Fig2}(a), for $U/J_x = 10$. In agreement with the known results \cite{Bukov2016}, we find that the thermalization is quick for both a ``high'' driving frequency $\omega/J_x \geq 20$ and for a ``low'' driving frequency $\omega/J_x \leq 10$.  For the intermediate values of $\omega$, the heating process is slow \cite{Bukov2016} and the total energy exhibits a slow exponential growth captured by
$
 Q(t = N T) \approx 1 - b \exp(-c\, t), t \gg 1.
$
An example of this behavior is given for $\omega/J_x = 15$ in Fig.~\ref{Fig:Fig2}(a). The heating process can also be monitored through   the particle-entanglement entropy $S_A$ as a function of time. In Fig.~\ref{Fig:Fig2}(b) for $N_p = 5$ and low driving frequency we find that this quantity quickly saturates to its maximal value. Indeed, for a thermal state at infinite temperature,  $S_A$ is given by 
\begin{equation}
S_A^{\text{max}} \approx \ln \binom{L_x \, L_y + N^A - 1}{N^A},
\label{eq:samax}
\end{equation}
marked by the horizontal (red) line in Fig.~\ref{Fig:Fig2}(b).
Except for the highest frequency considered ($\omega/J_x = 50$), we find that in the process of heating, the particle-entanglement gap of the initial state quickly closes (not shown in the plots).

Here we briefly discuss finite-size effects by comparing numerical results for the normalized total energy for $N_p = 4$, $N_p=5$ and $N_p = 6$. In line with the known results \cite{Lazarides2014, DAlessio2014, Bukov2015}, the ``high-frequency'' regime with low heating rates moves toward higher $\omega$ as the system size increases. However, we find that the estimates obtained in this section ($\omega/J_x\geq 20$ for the high, and $\omega/J_x\leq 10$ for the low-frequency regime, for $U/J_x = 10$) apply to all the three sizes $N_p = 4, 5, 6$, at least for the time scales that we consider.

 \begin{figure*}[!t]
\includegraphics[width=\linewidth]{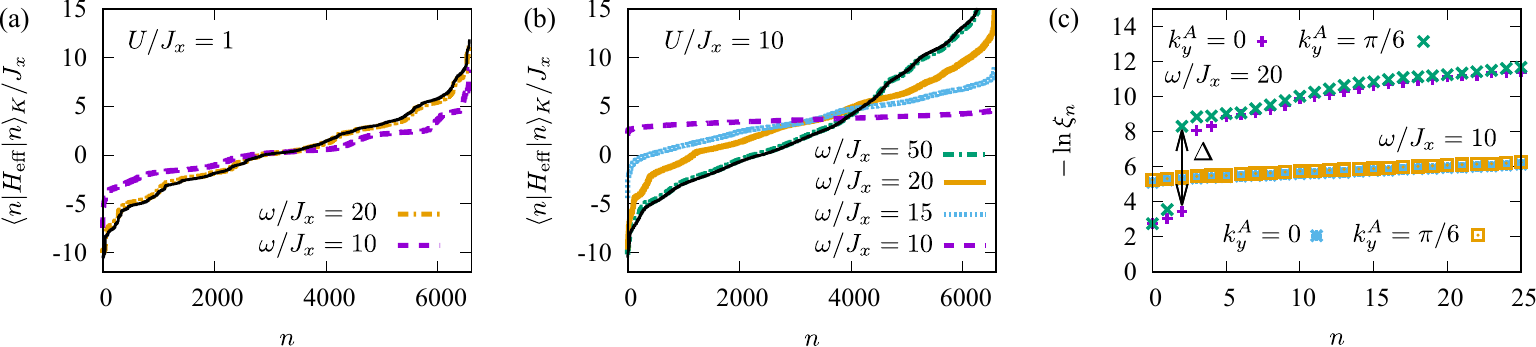}
\caption{Properties of the eigenstates $|n\rangle$ of the stroboscopic time-evolution operator $\hat U_F$, Eq.~(\ref{eq:UFn}), in the $k_x = 0, k_y = 0$ sector for $N_p = 4$.  Expectation values $\langle n| \hat H_{\text{eff}}|n\rangle_K$ defined in Eq.~\ref{eq:Hexpn} for (a) $U/J_x = 1$, $\omega/J_x = 10, 20$ and (b) $U/J_x = 10$, $\omega/J_x = 10, 15, 20, 50$. The black solid lines mark eigenenergies of $\hat H_{\text{eff}}$, Eq.~(\ref{eq:heff}). Note that in (b) we do not include few states from the top of the spectrum of $\hat H_{\text{eff}}$, Eq.~(\ref{eq:heff}) for clarity reasons. (c) The low-lying part of the particle-entanglement spectra $-\ln \xi_n$ of the incoherent superposition $\rho_F$,  Eq.~(\ref{eq:rhoF}), for $U/J_x = 10$, $\omega/J_x  = 20$ (crosses) and $\omega/J_x = 10$ (boxes).}
\label{Fig:Fig3}
\end{figure*}
\subsection{The stroboscopic time-evolution operator}
\label{subsec:Uf}

In order to better understand the limitations of the effective model, here we time evolve all relevant basis states for a single driving period $T = 2\pi/\omega$ and construct the stroboscopic time-evolution operator:
\begin{equation}
 \hat U_F \equiv \hat U(t_0 + T, t_0 = 0),
 \label{eq:UFn}
\end{equation}
such that $\hat U(N T + t_0) = \hat U_F^N$.
In the next step, for a system size $N_p = 4, (L_x, L_y) = (4, 8)$ we fully diagonalize this operator and inspect its eigenstates  $|n\rangle$. Following the described procedure, we obtain the long-time limit
\begin{equation}
 \lim_{N\rightarrow\infty}\langle \hat H_{\text{eff}}(N T)\rangle_K = \sum_n  |\langle n|\psi(t=0)\rangle|^2\langle n|\hat H_{\text{eff}}|n\rangle_K
 \label{eq:qN}
\end{equation}
where we define
\begin{equation}
\langle n|\hat H_{\text{eff}}|n\rangle_K = \langle n|e^{-i\hat K(t = 0)}\hat H_{\text{eff}}e^{i\hat K(t = 0)}|n\rangle.
\label{eq:Hexpn}
\end{equation}
Results for $ Q(t = N T)$ from Eq.~(\ref{eq:q}) obtained in this way are summarized in Fig.~\ref{Fig:Fig2}(c) where we make a comparison between the long-time energies for the case of hard-core bosons ($U\rightarrow \infty$) and soft-core bosons (finite values of $U$). The obtained results indicate that heating rates of hard-core bosons are closer to the case of $U/J_x = 1$ in comparison to $U/J_x = 10$, which is expected from the bandwidths shown in Fig.~\ref{Fig:Fig1}(a). Overall we observe that the ``high-frequency regime'' is wider for lower ratios $U/J_x$.

In Fig.~\ref{Fig:Fig3}, we make a comparison between the exact driven model captured by $ \hat U_F$ and $\hat H_{\text{eff}} $. In Figs.~\ref{Fig:Fig3}(a) and (b) we inspect the distribution of expectation values $\langle n |  \hat H_{\text{eff}}|n\rangle_K$. By comparing these values to the eigenenergies of the effective model, Eq.~(\ref{eq:heff}), we get an insight into the pertinence of the effective description \cite{DAlessio2014, Lazarides2014}. In particular, for an interacting system in the thermodynamic limit, the distribution is flat and the effective description is useless. We state again that we consider only small atomic samples. For this reason, it is expected that for high values of $\omega$, the full stroboscopic description nicely matches to the effective model values. Such an example is given in Fig.~\ref{Fig:Fig3}(a) for $U/J_x = 1$ and $\omega/J_x = 20$. As the value of $\omega$ gets lower the distribution becomes flatter, as can be seen in Fig.~\ref{Fig:Fig3}(b) for $U/J_x = 10$  by comparing  results for $\omega/J_x = 50$ and $\omega/J_x = 10$.

The intermediate regime of frequencies, e.~g.~$\omega/J_x = 20$ for $U/J_x=10$, is of the main experimental relevance \cite{Sun2018}. We now investigate whether the driven stroboscopic dynamics supports some Laughlin-like states, by calculating the PES of the mixture 
\begin{equation}
 \rho_{F} =\frac{1}{2}\big(|n_0(0, 0)\rangle \langle n_0(0, 0)| + |n_0(0, \pi)\rangle \langle n_0(0, \pi)|\big)
 \label{eq:rhoF}
\end{equation}
where $|n_0(k_x, k_y)\rangle$ is the state from the $k_x, k_y $ sector with the lowest expectation value $\langle n |  \hat H_{\text{eff}}|n\rangle_K$.
The results are presented in Fig.~\ref{Fig:Fig3}(c). We find that the states with a well defined gap and the Laughlin-like PES can be found down to $\omega/J_x\geq 15$ for $U/J_x = 1$, and down to $\omega/J_x\geq 20$ for $U/J_x=10$. Having established existence of these states for small samples of $N_p = 4$ particles, in the next section we discuss dynamical protocol which can be exploited to prepare these states.

\section{Slow ramp}
\label{sec:Slowramp}

The question about an optimal adiabatic protocol that can be used to prepare the Laughlin state in a cold-atom setup has gained lot of attention \cite{Popp2004, Sorensen, Motruk2017, He2017}. The situation becomes even more complex once the full driving process is taken into account. A general wisdom is that by starting from a topologically trivial state, the topological index of a thermodynamically large system can not be changed adiabatically. We consider a small atomic sample and follow the proposal of Ref.~\cite{He2017}. Our main contribution is that we extend this protocol to the case of the driven, interacting Bose-Hubbard model. 

\subsection{Model}
 Following results of Ref.~\cite{He2017}, we consider a slow ramp of the tunneling amplitude along $y$ direction, $J_y(t)$, as well as a slow ramp of the driving amplitude $\kappa(t)$. Namely, we start from a series of decoupled wires along the $x$ direction and start coupling them. More precisely, initial states are selected as the ground states of $\hat H_{\text{ini}}$
\begin{eqnarray}
  \hat H_{\text{ini}} &=& -J_x \sum_{m, n} \left(\hat  a_{m+1, n}^{\dagger} \hat a_{m, n} + \text{h.~c.~}\right)\nonumber\\
  &+& \frac{U}{2}\sum_{m, n}\hat n_{m, n}(\hat n_{m, n}-1).
  \label{eq:hpm}
 \end{eqnarray}
For the filling factors that we consider, the ground states of the $ \hat H_{\text{ini}}$ are simple non-interacting states with the ground state energy $E_{0, \text{ini}} = -2 J_x N_p$.
Out of the several degenerate ground states we select those where atoms occupy every second wire. There two such states and we label them as $|\psi_{+}\rangle$ (even wires occupied) and $|\psi_{-}\rangle$ (odd wires occupied). These states
have finite projections only onto the sectors $k_x = 0, k_y = 0$ and $k_x = 0, k_y = \pi$ of the driven model from Eq.~(\ref{eq:Hoft}).
Therefore we may expect the two initial states $|\psi_{\pm}(t = 0)\rangle$ to be transformed into the two Laughlin states during the ramp.

Having prepared the initial state, we slowly restore the tunneling amplitude along the $y$ direction, $J_y(t)$, and slowly ramp up the driving amplitude $\kappa(t)$. The time-evolution is governed by
 \begin{eqnarray}
  \hat H_{\text{sr}}(t) &=& -J_x \sum_{m, n} \left(\hat  a_{m+1, n}^{\dagger} \hat a_{m, n} + \text{h.~c.~}\right)\nonumber\\
  &-& J_y(t) \sum_{m, n}  \left(e^{i \omega t}\hat a_{m, n+1}^{\dagger} \hat a_{m, n}  + \text{h.~c.~}\right)\nonumber\\
  &+& \frac{\kappa(t)}{2}\sum_{m, n} \sin\left(\omega t - (m + n -1/2)\, \phi\right) \hat n_{m,n}\nonumber\\
  &+& \frac{U}{2}\sum_{m, n}\hat n_{m, n}(\hat n_{m, n}-1),
  \label{eq:sr}
 \end{eqnarray}
 where
 $J_y(t) = J_y \tanh(\eta \,t), \, \kappa(t) = \kappa \tanh(\eta\, t)$, $\eta$ being the ramping rate. In the long-time limit, we recover the original Hamiltonian from Eq.~(\ref{eq:Hoft}).
 During the ensuing time evolution we construct the mixture
\begin{equation}
 \rho(t) = \frac{1}{2}\big(|\psi_{+}(t)\rangle \langle \psi_{+}(t)| + |\psi_{-}(t)\rangle \langle \psi_{-}(t)| \big).
 \label{eq:rhopm}
\end{equation}
We monitor stroboscopically the energy expectation value 
\begin{equation}
E(t)=\text{tr}\left(\rho(t)\hat H_{\text{eff}}\right)
\label{eq:heffexppm}
\end{equation}
and the PES of $\rho(t)$.

 \begin{figure*}[!t]
\includegraphics[width=\linewidth]{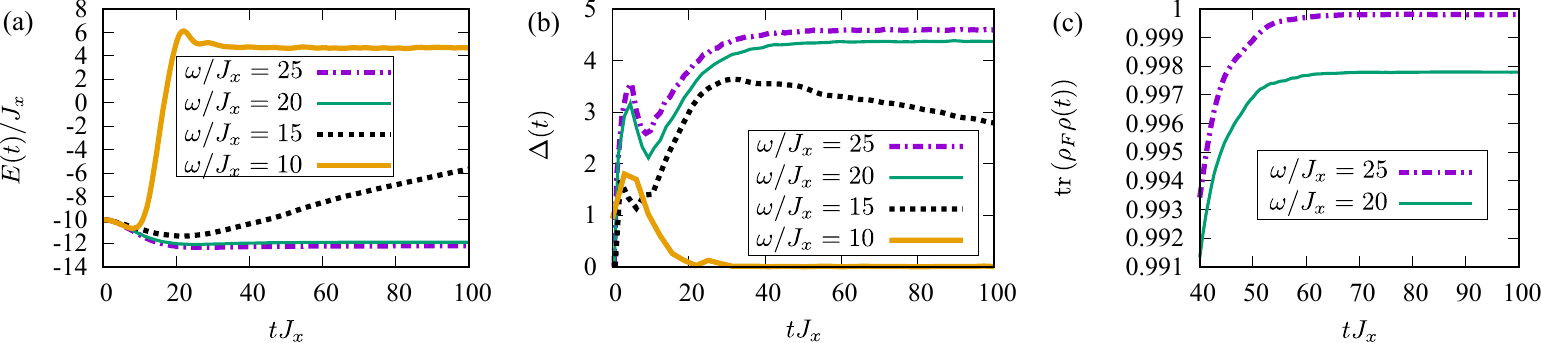}
\caption{(a) The expectation value $E(t)$ defined in Eq.~(\ref{eq:heffexppm}) and (b) the particle-entanglement gap $\Delta(t)$ of $\rho(t)$, Eq.~(\ref{eq:rhopm}), during the time evolution governed by Eq.~(\ref{eq:sr}) for several driving frequencies $\omega/J_x = 25,\, 20,\, 15,\, 10$. Parameters: $N_p = 5, U/J_x = 10, \eta/J_x = 0.05$.  (c) The overlap $\text{tr}\,\left(\rho(t) \rho_F\right)$ of the time evolved state with the target eigenstates of $\hat U_F$ for $\omega/J_x = 25,\, 20$.   Parameters: $N_p = 4, U/J_x = 10, \eta/J_x = 0.05$.  
}
\label{Fig:Fig4}
\end{figure*}

\subsection{Results}

In Fig.~\ref{Fig:Fig4}(a) we present the energy expectation value from Eq.~(\ref{eq:heffexppm}) for $U/J_x = 10$ and several driving frequencies $\omega/J_x = 25,\, 20,\, 15,\, 10$. Our numerical results indicate that ramps with the rates up to $\eta/J_x \sim 0.1$ work reasonably well. Slower ramps give better results, but are less practical \cite{He2017}. By construction, the initial state is a non-interacting state with particles delocalized along the $x$ direction and therefore the initial energy is $E(t = 0) = -2 \, N_p\, J_x$.
During the ramp with the rate $\eta/J_x = 0.05$, for the regime of high driving frequencies, down to approximately $\omega/J_x = 20$, we find that  the energy initially decreases and reaches an almost constant value at around $t J_x \sim 20$. On the other hand, for $\omega/J_x = 15$, the system slowly heats up during the ramping process, and for $\omega/J_x = 10$ the system quickly reaches the infinite-temperature state.

One of our main results is summarized in Fig.~\ref{Fig:Fig4}(b), where we plot the particle-entanglement gap of $\rho(t)$, from Eq.~(\ref{eq:rhopm}), as a function of time. In the high-frequency regime $\omega/J_x \geq 20$, starting around $t J_x \sim 20$ we find a persistant particle-entanglement gap, marking the onset of a topologically non-trivial state. It is even more interesting, that even for $\omega/J_x \sim 15$, the state seems to exhibit a finite gap on intermediate time scales. This is not the case for $\omega/J_x \leq 10$, where the gap quickly vanishes. In Fig.~\ref{Fig:Fig4}(c), we present the value of the overlap $\text{tr}\,\left(\rho(t) \rho_F\right)$, of the time-evolved mixed state with the relevant state from Eq.~(\ref{eq:rhoF}) for $N_p = 4$. Clearly, the slow ramp of the type given in Eq.~(\ref{eq:sr}), allows for the preparation of the relevant eigenstates of $\hat U_F$ with high fidelity (better than 1$\%$).
 
\begin{figure}[!t]
\includegraphics[width=\linewidth]{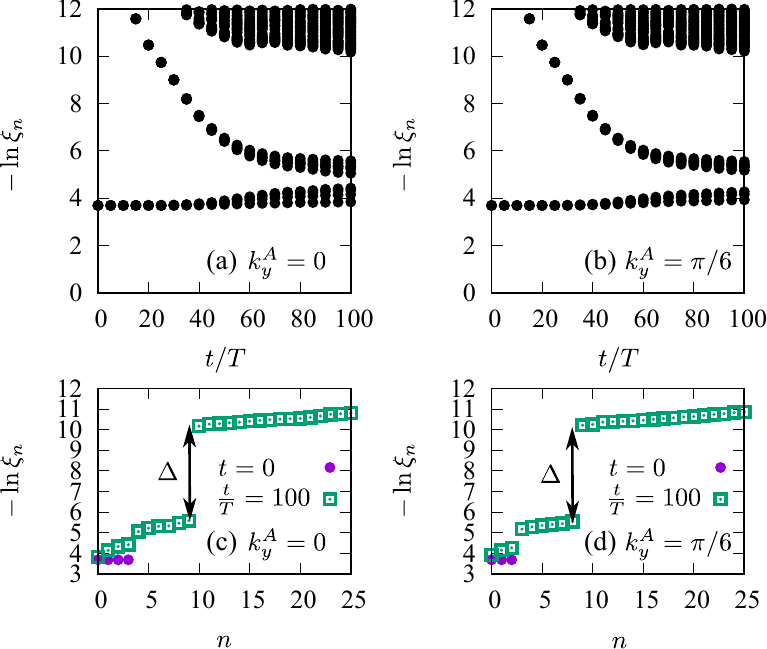}
\caption{The low-lying part of the particle-entanglement spectra $-\ln \xi_n$ of $\rho(t)$, Eq.~\ref{eq:rhopm}, during the time evolution governed by Eq.~(\ref{eq:sr}) in the (a) $k_y^A = 0$, and (b) $k_y^A = \pi/6$ momentum sectors. The low-lying part of the PES in the sectors (c) $k_y^A = 0$, and (d) $k_y^A = \pi/6$, at two instances of time $t = 0$ and $t/T = 100$. Parameters: $N_p = 6, U/J_x = 5, \omega/J_x = 15, \eta/J_x = 0.05$.
}
\label{Fig:Fig5}
\end{figure}

In Figs.~\ref{Fig:Fig5} (a) and (b) we show the time evolution of the PES in the two momentum sectors $k_y^A = 0$ and $k_y^A= \pi/6$ for $N_p = 6$, $U/J_x = 5$ and $\eta/J_x =0.05$. The PES of the initial state is easy to understand. As the $L_y/2$ wires are occupied by single atoms, the reduced density matrix is proportional to the identity matrix with the proportionality factor yielding $-\ln \xi_n = \ln \left(2\binom{Ly/2}{N_A}\right)\approx 3.69$. During the ramp we find that additional modes in PES are gaining weight and moving down in the spectrum.
Finally,  the state $\rho(t)$ reached around $t \approx 50 T$ exhibits a well defined gap and the correct counting of the low-lying modes: there are $10$ low-lying modes for $k_y^A = 0$ and $9$ low-lying modes for $k_y^A = \pi/6$, see  Figs.~\ref{Fig:Fig5} (c) and (d), see also Table \ref{Tab:Tab1}.

\begin{figure}[!t]
\includegraphics[width=\linewidth]{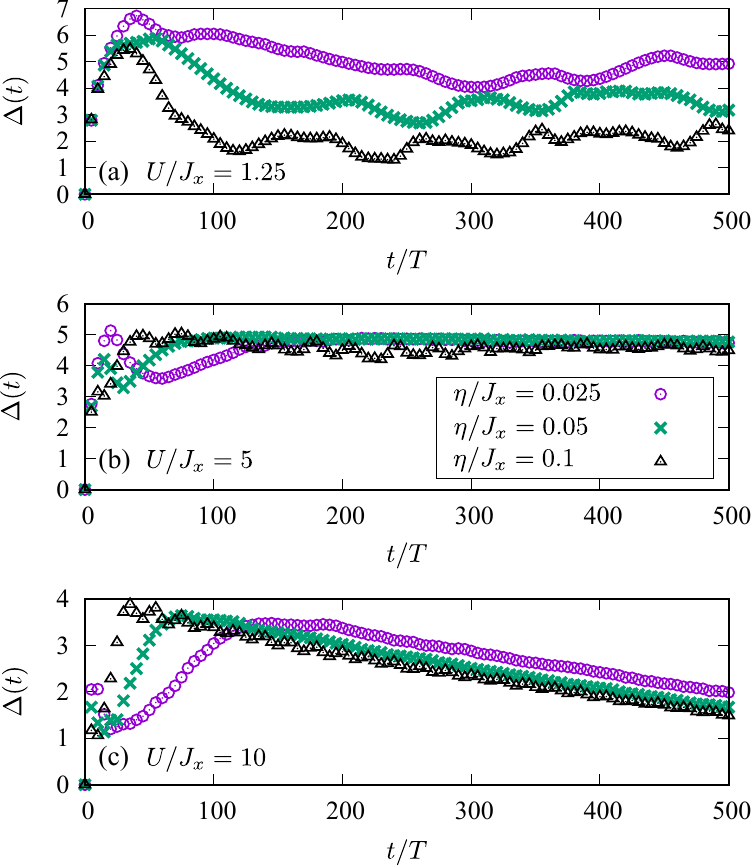}
\caption{The particle-entanglement gap $\Delta(t)$ as a function of time during the time evolution governed by Eq.~(\ref{eq:sr}), for several interaction strengths (a) $U/J_x = 1.25$, (b) $U/J_x = 5$, and (c) $U/J_x = 10$, and several ramping rates $\eta/J_x = 0.025, 0.05, 0.1$. Other parameters: $N_p = 5, \omega/J_x = 15$. 
}
\label{Fig:Fig6}
\end{figure}

In Fig.~\ref{Fig:Fig6}  we discuss a satisfactory range of ramping rates $\eta$ for a given interaction strength $U$ and a given driving frequency $\omega$ that we fix at $\omega/J_x = 15$. The obtained numerical results suggest that at weaker interaction strengths $U/J_x \leq 2$, slower ramping rates are needed. One way to explain this behavior is by using the effective model and arguing that the gap protecting the Laughlin state is smaller at weaker $U$.
On the other hand, for stronger interaction strengths $U/J_x \geq 8$ the particle-entanglement gap closes at later stages as the heating process becomes dominant. Finally, in the intermediate range $U/J_x \sim 5$, faster ramps with $\eta/J_x = 0.1$ lead to the sought-after state $\rho(t)$ from Eq.~\ref{eq:rhopm}, with persistant features in the PES up to $t = 500 T$.
These results indicate that when optimizing the ramping protocol in an actual experiment, there will be a trade off between the unfavorable heating and a faster ramping into the desired state, as both of these processes are promoted by interactions.
\section{Conclusions}

The technique of Floquet engineering has been successfully exploited for the implementation of synthetic magnetic fields in driven optical lattices. Following up on these achievements and on a long-standing pursuit for the FQH states in cold-atom setups,
in this paper we have addressed possible realization of the bosonic Laughlin state in a small atomic sample in a periodically driven optical lattice. While a thermodynamically large interacting system generally heats up into an infinite-temperature state under driving, the heating process can be controlled to some extent in a few-particle system.

We have assumed a realistic driving protocol and finite on-site interactions, and we have identified the FQH state based on analysis of its particle-entanglement spectra. Results of our numerical simulations show that the stroboscopic dynamics of $N_p = 4, 5, 6$ particles supports the topological $\nu=1/2$ Laughlin state down to $\omega/J_x=20$ for $U/J_x = 10$, and down to $\omega/J_x = 15$ for $U/J_x = 1$, for the driving amplitude $\kappa/\omega = 0.5$. These results are in reasonable agreement with the recent estimates of the optimal heating times \cite{Sun2018} that take into account the contribution of the higher bands of the underlying optical lattice. In addition, we have investigated slow ramping of the driving term and found that it allows for the preparation of the Laughlin state on experimentally realistic time scales of the order of $20 \,\hbar/J_x$, where $\hbar/J_x$ is the  tunneling time. Interestingly, we find that some topological features persist during an intermediate stage even in the regime where the system exhibits a slow transition into the infinite-temperature state (e.~g.~$\omega/J_x=15$ for $U/J_x = 10$).

A highly relevant question that we have not tackled and that we postpone to future investigation, concerns suitable experimental probes of topological features. The recent progress in the field has led to the development of several detection protocols specially suited for the cold-atom systems \cite{Price2012, Dauphin2013, Tran2017, Hudomal2018, Repellin2018, Raciunas2018}. For the type of systems considered in this paper, the most promising are results of the recent study \cite{Raciunas2018} showing that fractional excitations can be probed even in small systems of several bosons.

\section{Acknowledgments}
 
 This work was supported by the Ministry of Education, Science,  and  Technological  Development  of  the  Republic of Serbia under Project ON171017. NR was supported by the grant ANR TNSTRONG No.~ANR-16-CE30-0025 and
ANR TopO No.~ANR-17-CE30-0013-01.  Numerical  simulations  were  performed  on  the  PARADOX  supercomputing facility at the Scientific Computing Laboratory of the Institute of Physics Belgrade.  
The authors would also like to acknowledge the contribution of the COST Action CA16221.‌

  \onecolumngrid
 \appendix*

 \section{Driven optical lattices}
 In this appendix we review the derivation of the model given in Eq.~(\ref{eq:Hoft}).
 The system is described by 
 \begin{equation}
\hat H_{\text{lab}}(t) = \hat H_{BH} + \hat H_{\text{drive}}(t) + \omega\, \hat V,
 \end{equation}
 where we start with the Bose-Hubbard model
 \begin{equation}
  \hat H_{BH} = -J_x \sum_{m, n} \left(\hat a_{m+1, n}^{\dagger} \hat a_{m, n} + \mathrm{h.~c.~}\right)-J_y \sum_{m, n} \left(\hat a_{m, n+1}^{\dagger} \hat a_{m, n} + \mathrm{h.~c.~}\right) + \frac{U}{2}\sum_{m, n}\hat n_{m, n}(\hat n_{m, n}-1) ,
 \end{equation}
 and we introduce an offset $\omega \hat V$ 
 \begin{equation}
 \hat V= \sum_{m, n} n\, \hat n_{m,n}.
 \end{equation}
 This shifted Bose-Hubbard model is exposed to a suitable resonant driving scheme:
\begin{equation}
 \hat H_{\text{drive}}(t) = \frac{\kappa}{2}\sum_{m, n} \sin\left(\omega t -\phi_{m, n} + \frac{\phi}{2}\right) \hat n_{m,n}, \quad \phi_{m, n} = (m + n)\,\phi.
 \label{eq:drive}
\end{equation}
We assume periodic boundary conditions compatible with the driving term (\ref{eq:drive}) in the lab frame.  To this purpose we use vectors $\mathbf{R}_1 = 4\, \mathbf{e}_x$ and $\mathbf{R}_2 =- \mathbf{e}_x+\mathbf{e}_y$ as presented in Fig.~\ref{Fig:Fig0}. 
For simplicity, we work in the rotating frame 
\begin{equation}
 |\psi_{\text{rot}}(t)\rangle = e^{i \omega t \hat V } |\psi_{\text{lab}}(t)\rangle
\end{equation}
and derive the Schr\"odinger equation
\begin{equation}
 i \frac{d |\psi_{\text{rot}}(t)\rangle}{dt} = \hat H_{\text{rot}}(t)|\psi_{\text{rot}}(t)\rangle,
 \label{eq:rotframe}
\end{equation}
where 
\begin{equation}
\hat H_{\text{rot}}(t) =  \left(e^{i \omega t \hat V }\hat H_{\text{lab}}(t)e^{-i \omega t \hat V }-\omega \hat V\right).
\end{equation}
Now we calculate $ \hat H_{\text{rot}}(t)$ explicitly. The only nontrivial action of this rotation on $\hat H_{\text{lab}}$ comes from the  nearest-neighbor hopping along $y$ direction. Indeed, we have
\begin{equation}
 e^{i \omega t \hat V }\hat a_{m, n}^{\dagger} \hat a_{m, n'}e^{-i \omega t \hat V } = e^{i \omega t (n-n')} \hat a_{m, n}^{\dagger} \hat a_{m, n'}.
\end{equation}
In total we obtain
\begin{eqnarray}
 \hat H_{\text{rot}}(t) &=& -J_x \sum_{m, n} \left(\hat a_{m+1, n}^{\dagger}  \hat a_{m, n} + \mathrm{h.~c.~}\right) + \frac{U}{2}\sum_{m, n}\hat n_{m, n}(\hat n_{m, n}-1)\nonumber\\ &+& e^{i \omega t} \hat H_1 + e^{-i \omega t}\hat H_{-1}+e^{-i \omega t(L_y - 1)}  \hat H_{L_y-1}+e^{i \omega t(L_y - 1)}  \hat H_{-L_y+1},
 \label{eq:hprimeoft}
 \end{eqnarray}
 with
 \begin{eqnarray}
  \hat H_1 &=&-J_y \sum_{m, n}^{\text{OBC}} \left(\hat a_{m, n+1}^{\dagger} \hat a_{m, n}-\frac{i}{4} \kappa e^{i (-\phi_{m, n}+  \frac{\phi}{2})}\hat n_{m,n}\right),\quad \hat H_{-1} = \hat H_1^{\dagger},\\
 \hat H_{-L_y+1} &=&-J_y \sum_{m} \hat a_{m, 0}^{\dagger} \hat a_{m - L_y, L_y-1},\quad \hat H_{L_y-1} = \hat H_{-L_y+1}^{\dagger}.
 \end{eqnarray}
In the terms $\hat H_{-L_y+1}$ and $\hat H_{L_y-1}$ we take into account periodic boundary conditions along the direction parallel to ${\bf R_2}$ as imposed in the lab frame. 
In order to limit the complexity of the numerical calculation, we keep translational invariance and impose the periodic boundary conditions in both directions in the rotating frame. 
This implies that we will neglect ``phasors'' $e^{-i \omega t(L_y - 1)}$ and $e^{i \omega t(L_y - 1)} $. Under these assumptions, we can recast Eq.~(\ref{eq:hprimeoft}) into the time-dependent Hamiltonian given in Eq.~(\ref{eq:Hoft}).
In practice, this would require engineering additional non-trivial terms in the lab frame. 

The leading order of the kick operator is given by
\begin{equation}
 \hat K(t = 0) \approx-\frac{\kappa}{2  \omega} \sum_{m, n}  \cos (\phi_{m,n}-\phi/2) \hat n_{m,n}.
 \label{eq:kick}
\end{equation} 

\twocolumngrid

\end{document}